\documentclass[preprint,aps,prd,showpacs,superscriptaddress,nofootinbib]{revtex4}

\usepackage{amsfonts}
\usepackage{mathrsfs}
\usepackage{graphicx}
\usepackage{amsmath}
\usepackage{amssymb}
\usepackage{bm}
\usepackage{bbm}
\usepackage{color}
\def\OMIT#1{}

\newcommand{\nn}{\nonumber}

\newcommand{\beq}{\begin{equation}}
\newcommand{\eeq}{\end{equation}}
\newcommand{\bqa}{\begin{eqnarray}}
\newcommand{\eqa}{\end{eqnarray}}

\begin{document}

\title{ Higher Order Corrections to the Cross Section of
$e^+e^- \to {\rm quarkonium} + \gamma$ }

\author{Wen-Long Sang}
\affiliation{Key Laboratory of Frontiers in Theoretical
             Physics,Institute of Theoretical Physics, Academia Sinica,
             Beijing 100190, China \vspace*{1.5cm}}
\author{Yu-Qi Chen}
\affiliation{Key Laboratory of Frontiers in Theoretical
             Physics,Institute of Theoretical Physics, Academia Sinica,
             Beijing 100190, China \vspace*{1.5cm}}
\begin{abstract}
The cross sections of $e^+e^- $ to $S-$wave and $P-$wave quarkonia
with $C$-parity even associated with a photon
are analyzed in the framework of non-relativistic Quantum
Chromodynamics(NRQCD) factorization formulism. The short-distance
coefficients are analytically determined up to the next-to-leading order(NLO) QCD
radiative corrections to $S-$wave and $P-$wave quarkonium production and NLO relativistic
correction to $\eta_{c}$ production. We also invoke the analytical expressions
to estimate the cross sections. Our numerical results indicate
that both the QCD and the relativistic corrections are considerable.
\end{abstract}
% 12.38.-t   Quantum chromodynamics
% 12.39.St  Factorization
% 13.20.Gd  Decays of J/psi, Upsilon, and other quarkonia
% 13.66.Bc Hadron production in e?e+ interactions
% 14.40.Gx   Mesons with S=C=B=0, mass > 2.5 GeV (including quarkonia)

%\keywords{}

\maketitle

%--------------------------------------------------------------------
%%%%%%%%%%%%%%%%%%%%%%%%%%%%%%%%%%%%%%%%%%%%%%%%%%%%%%%%%%%%%%%%%%%%%%%%%%%%%%
\section{Introduction}\label{sec:introduction}
%%%%%%%%%%%%%%%%%%%%%%%%%%%%%%%%%%%%%%%%%%%%%%%%%%%%%%%%%%%%%%%%%%%%%%%%%%%%%%
Non-relativistic quantum chromodynamics(NRQCD) factorization
formalism\cite{Bodwin:1994jh} is an useful tool for analyzing the
inclusive production  of the heavy quarkonium. According to it, the
cross section is expressed as a sum of products of short-distance
coefficients and NRQCD matix elements. The short-distance
coefficients can be calculated as perturbation series in coupling
constant $\alpha_s$ at the scale of the heavy quark mass. The matrix
elements scale in a definite way with the typical relative velocity
$v$ of heavy quark in the quarkonium state. Thus the production
cross sections can be expressed as  double expansions both  in
$\alpha_s$ and in $v$  to any desired order.

Recently, the cross section of the exclusive process $e^+e^-\to
H+\gamma$ at the center-of-momentum(CM) energy $\sqrt{s}=10.58$GeV
was computed by Chung, Lee, and Yu\cite{jungil Lee}, where $H$
represents a $S-$wave or $P-$wave heavy quarkonium state with
charge-conjugation parity $C=+1$, to the leading order(LO) both in
$\alpha_s$ and in $v$ in the NRQCD factorization formalism. Their
predicted cross sections are about $82{\rm fb}$, $49{\rm fb}$, $1{\rm fb}$, $14{\rm fb}$ and $5{\rm fb}$
for $\eta_c$, $\eta_c(2S)$, $\chi_{c0}$, $\chi_{c1}$ and $\chi_{c2}$ production, respectively.  Some
early calculations of similar processes like $Z^0 \to H + \gamma$
were presented in \cite{Guberina:1980dc} in the color-singlet model.

%In experiments, the heavy quarkonium can be
%reconstructed either by measuring the recoiled photon or by
%measuring the decay products of quarkouium.

These relatively large cross sections make the processes be measurable
at the B factories. To have more precise predictions, it is
necessary to carry out the calculations to higher order both in
$\alpha_s$ and in $v$. One may expect that those correction terms
are considerable.  Especially, for charmonium production, both
$\alpha_s$ and $v^2$ are not small.
 In this paper, in the framework of NRQCD factorization
formalism, we compute  the short-distance coefficients of the
leading NRQCD matrix element to the NLO QCD corrections both for
$S-$wave and $P-$wave heavy quarkonium production. For the $\eta_c$,
we also calculate the tree level short-distance coefficient of the $v^2$
relativistic correction term. We will not perform the calculation for the NLO relativistic corrections to $\eta_{c}(2S)$ and $P-$wave quarkonium production
since the NLO NRQCD matrix elements for these states
are still not determined preciously by fitting data. The NLO QCD correction to the $e^+e^-\to \eta_c+\gamma$ was
calculated first in \cite{Shifman:1980dk}. However, we found that
their original expressions given in Eqs.~(23), (25) in their paper
were incorrect. Probably there were some typos in their equations.

The NRQCD factorization formulism  separates  the short-distance
 effects happened at energy
scale $m$ or higher, where $m$ is the heavy quark mass, and the
long-distance effects happened at long-distance $1/mv$ or larger
which describes the formation of the heavy quarkonium. The
short-distance effects are described by the short-distance
coefficients while the long-distance ones are described by the NRQCD
long-distance matrix elements. In the process $e^+e^-\to H +\gamma$,
the emission of the photon can be either short-distance effects or
long-distance ones depending on the emitted photon being  hard  or
soft comparing to the quark mass $m$. Thus when the NRQCD
factorization approach is used to analyze the process of
$e^+e^-\to H+\gamma$ at the B factories, the factorization formulism
for the cross sections
% is applied to analyze to processes
takes different forms in the cases of $H$ being  $c\bar{c}$ or
$b\bar{b}$ state. In the former case, the emitted photon is so hard
 that the short-distance process is $e^+e^-\to c\bar{c} +\gamma$.
However, in the later one, the energy of the emitted photon is much
less than the $b$ quark mass so that the emission of the photon  is
a long-distance effect and should be described by the long-distance
matrix as an electric $E1$ or a magnetic $M1$ transition.

The rest of this article is organized as follows. In Section~II, we
apply the NRQCD factorization formulism to analyze $e^+e^-\to H
+\gamma$ at the B factories where we distinguish two different cases
of $c\bar{c}$ or $b\bar{b}$, corresponding to the emitted photon
being hard or soft. In Section~III, we determine the short-distance
coefficient for the cross section of the process $e^+e^-\to
\eta_c+\gamma$ up to the NLO radiative QCD correction for the LO
matrix element and the LO short-distance coefficient for the
relativistic correction term. We also discuss various limitations
of those short-distance coefficients. In Section~IV, we determine
the short-distance coefficients for the cross sections of the
processes of $e^+e^-\to \chi_{cJ}+\gamma$ up to  the NLO QCD radiative
corrections and discuss various limitations. Section~V contributes
to discussions of the factorization form for the processes of
$e^+e^-\to \eta_b( \chi_{bJ})+\gamma$ with the emitted photon being
soft.   In Section~VI, we apply the obtained formulae
to carry out numerical estimations for the cross sections of those
processes. We also summarize our results in this section. We present
the analytical  NLO short-distance coefficients of the leading order
NRQCD matrix elements in $v$ for the cross sections of $e^+e^-\to
\eta_c+\gamma$ and $e^+e^-\to \chi_{cJ}+\gamma$ in Appendix A and
Appendix B, respectively.

%%%%%%%%%%%%%%%%%%%%%%%%%%%%%%%%%%%%%%%%%%%%%%%%%%%%%%%%%%%%%%%%%%%%%%%%%%%%%%
\section{ NRQCD factorization formulism for $e^+e^- \to H +\gamma$}
 \label{sec:2}
%%%%%%%%%%%%%%%%%%%%%%%%%%%%%%%%%%%%%%%%%%%%%%%%%%%%%%%%%%%%%%%%%%%%%%%%%%%%%%

In the process of $e^+e^- \to H+\gamma$,  where $H$ represents a
$S-$wave or $P-$wave heavy quarkonium state with C-parity even, an almost on-shell and near threshold $Q\bar{Q}$ pair
is created at the short-distance scale of $1/m$ or smaller, then constitutes a heavy quarkonium $H$ that happened at the distance
scale of $1/mv$. The production cross
section can be  analyzed in the NRQCD factorization
formulism{\cite{Bodwin:1994jh}} as double expansions both in velocity
$v$ and in coupling $\alpha_s$. The factorization takes two different forms
in two different cases corresponding to the emitted photon
being hard and soft.

For charmonium production at the B factories via process $e^+e^- \to
\,{\rm charmonium }\,+\gamma$, the momentum of the emitted photon
satisfies $|k| \geq m$. The photon is so hard that
% According to the NRQCD factorization formula,
 the short-distance effects arise from the process $e^+e^- \to
c\bar{c}+\gamma$. We assign  ${\cal K}_n$
with $n$ in the collection of ${^1S_0}$, $^3P_{0}$, $^3P_{1}$ and $^3P_{2}
$ as follows\cite{Bodwin:1994jh}:
\begin{eqnarray}\label{NRQCDoperator}
{\cal K}_{^1S_0}&=&1,\nonumber\\
{\cal K}_{^3P_{0}}&=&\frac{1}{\sqrt{3}}(-\frac{i}{2}\tensor{{\bm
D}}\cdot\sigma),\nonumber\\
{\cal K}_{^3P_1}&=&\frac{1}{\sqrt{2}}(-\frac{i}{2}\tensor{{\bm
D}}\times\sigma),\nonumber\\
{\cal K}_{^3P_2}&=&-\frac{i}{2}\tensor{ D^{(i}}\sigma^{j)},
\end{eqnarray}

The NRQCD factorization formulism for the cross section of $e^+e^- \to
H+\gamma_{hard}$ ($H=\eta_c$, $\eta_c(2S)$, $\chi_{cJ}$) takes the following uniform form:
%the cross section of the exclusive $\eta_c(\eta_b)+\gamma$ from
%$e^+e^-$ annihilation can be written as
%
\begin{eqnarray}\label{eq:formular-hard}
\sigma %[e^+e^- \to H +\gamma]
 &=&\frac{F_1({n})}{m_c^2}
\langle0|\chi^{\dagger}{\cal K}_n \psi|H\rangle\langle H|
\psi^{\dagger}{\cal K}_n \chi|0\rangle
\nonumber\\
&+&\frac{G_1({n})}{m_c^4}{\rm Re}\langle0|\chi^{\dagger}{\cal K}_n
\psi|H\rangle\langle H| \psi^{\dagger}{\cal K}_n
(-\frac{i}{2}\tensor{\bf{D}})^2\chi|0\rangle ,
\end{eqnarray}
where $\langle0|\chi^{\dagger}{\cal
K}_n \psi|H\rangle \langle H|\psi^{\dagger}{\cal K}_n
\chi|0\rangle$ and ${\rm Re}\langle0|\chi^{\dagger}{\cal K}_n \psi|
H\rangle\langle H| \psi^{\dagger}{\cal K}_n
(-\frac{i}{2}\tensor{\bf{D}})^2\chi|0\rangle$ are NRQCD matrix
elements and $F_1(n)$ and $G_1({n})$ are the short-distance
coefficients corresponding to the matrix elements. The leading
contributions to the matrix elements arise from these states, which possess the same quantum numbers as
$n$. In (\ref{eq:formular-hard}), the NRQCD matrix element in the
second term is suppressed by $v^2$ compared to that in the first
term. The second term is usually called as the relativistic
correction term.
 The short-distance coefficients can be expanded as power series of $
\alpha_s $ at energy scale of $m$ or higher. They can be determined
by matching the cross section of a free quark pair production
process $e^+(p_{e1}) +e^-(p_{e2}) \to c(p_1)\bar{c}(p_2) +
\gamma(k)$ with the color-singlet on-shell $c\bar{c}$  pair near the
threshold. Given the total and the relative momenta of the
$c\bar{c}$ pair are $P$ and $2q$, respectively, we then have
\begin{eqnarray}
p_1=\frac{1}{2}P+q, ~~~ p_2=\frac{1}{2}P-q,~~~p_1^2=p_2^2=m^2,~~~
P\cdot q =0\;. \nonumber
\end{eqnarray}
Then the S-matrix element for this process is given by
\begin{eqnarray}{\label{treelevelamp}}
-i{\cal M}=-i\frac{e}{s}L_\mu{\cal A}^{\mu\nu}\varepsilon_\nu^*,
\end{eqnarray}
where $e$ is the electromagnetic coupling, $\varepsilon$ is the
polarization four-vector of the photon and $s$ is the square of the
center-of-mass(CM) frame energy. The leptonic current $L_\mu$ in
(\ref{treelevelamp}) is defined by
\begin{eqnarray}
L_\mu=\bar{v}(p_{e2})\gamma_\mu u(p_{e1}),
\end{eqnarray}
and the amplitude ${\cal A}^{\mu\nu}$ is given by
\begin{eqnarray}
{\cal A}^{\mu\nu}={\rm Tr}\,[\,{\tilde A}^{\mu\nu}\Pi_m\,],
\label{aaa}
\end{eqnarray}
where $\Pi_m$ is the projector operator of spin-single ($m=0$) or
spin-triplet ($m=1$) state and ${\tilde{\cal A}}^{\mu\nu}$ is the
amplitude of $e^+ +e^- \to c\bar{c} + \gamma$ with the
wave-functions of the external quark lines  removed. For the
spin-singlet state, $\Pi_0$ reads\cite{Bodwin:2002hg}
\begin{eqnarray}{\label{projector-1}}
\Pi_0&=&\frac{(\not\!P /2-\not\!q-m)\gamma_5(\not\!P+2E)(\not\!P
/2+\not\!q+m)}{4\sqrt{6}E(E+m)}\,
\end{eqnarray}
and for the spin-triplet state with spin-polarization vector
$\epsilon$, $\Pi_1$ reads\cite{Bodwin:2002hg}
\begin{eqnarray}{\label{projector-2}}
\Pi_1 &=-&\frac{(\not\!P /2-\not\!q-m)\not\!\epsilon^\star \,
(\not\!P+2E)(\not\!P /2+\not\!q+m)}{4\sqrt{6}E(E+m)}\,,
\end{eqnarray}
where $E=\sqrt{m^2+{\bf q}^2}$.

%%%%%%%%%%%%%%%%%%%%%%%%%%%%%%%%%%%%%%%%%%%%%%%%%%%%%%%%%%%%%%%%%%%%%%%%%%%%%%
\begin{figure}[!hbp]
\centerline{
\includegraphics[width=15.0cm]{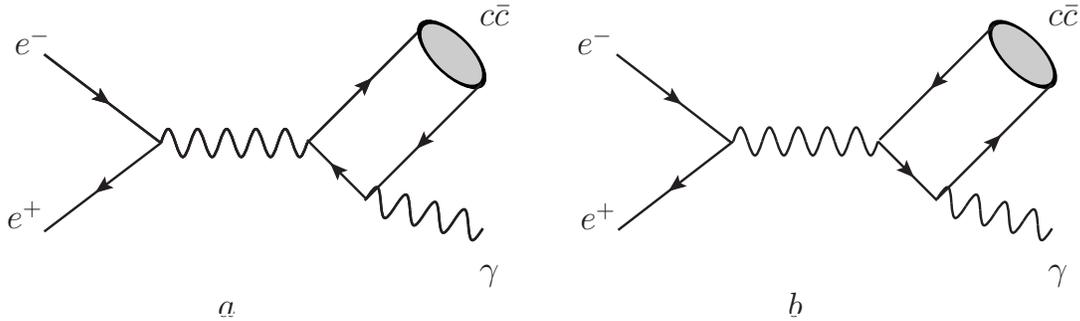}
} \caption{The tree level Feynman diagrams for   $e^+e^- \to c\bar{c}
+\gamma$. } \label{fig1}
\end{figure}

At tree level, there are two Feynman diagrams contributing to the
process as illustrated in Fig.~\ref{fig1}. ${\tilde{\cal A}}^{\mu\nu}$
is written as
\begin{eqnarray}\label{A-mn-tilde}
{\tilde{\cal
A}}^{\mu\nu}=e_Q^2e^2\left[\gamma^\mu\frac{-\not\!p_2-\not\!k+m}
{(p_2+k)^2-m^2}\gamma^\nu+\gamma^\nu\frac{\not\!p_1
+\not\!k+m}{(p_1+k)^2-m^2}\gamma^\mu\right],
\end{eqnarray}
where $e_Q$ is the electric charge of the heavy quark.

By expanding the amplitude in terms of $q$, the contributions
arising from $S-$wave and $P-$wave can be picked out. The
cross section in the QCD can then be evaluated by
squaring the amplitude and integrating over the phase space. The
cross section of the same process can also be evaluated in the
NRQCD factorization formulism with the hadron in the NRQCD matrix elements
 replaced with the free quark pair being of the same quantum number as $H$. The short-distance coefficients $F_1({}n)$ and
$G_1(n)$ can then be inferred from equaling the cross section calculated both in the QCD
and in the NRQCD factorization formulism. In our paper, we determine the $F_1({^1S_0})$ and
$F_1({^3P_J})$ to the NLO QCD corrections in $\alpha_s$ and
$G_1({^1S_0})$ to the LO in $\alpha_s$.

On the other hand,  for bottomonium production at the B factories via
process $e^+e^- \to \,{\rm bottomium}\,+\gamma$, the momentum of the
emitted photon yields $|k| \ll m$. The photon is so soft that the
wave-length of the emitted photon is compatible with the size of the
bottomonium. Hence the emission process is sensitive to the
long-distance effects. Its effects should be absorbed into the long-distance NRQCD
matrix elements. The short-distance effects arise from the
contributions of the process $e^+e^- \to b\bar{b}$ with $b\bar{b}$
in a spin-triplet and a color-singlet state. So, the NRQCD factorization
formulism for the cross section of $e^+e^- \to H+\gamma_{soft}$ takes
the following form:
\begin{eqnarray}\label{eq:formular-soft}
%[e^+e^- \to H+\gamma]
 \sigma&=&\frac{F_2({}^3S_1)}{m^2}
\langle0|\chi^{\dagger}\sigma^i\psi|H+\gamma \rangle\langle
H+\gamma| \psi^{\dagger}\sigma^i\chi|0\rangle
\nonumber\\
&+&\frac{G_2({}^3 S_1)}{m^4}{\rm
Re}\langle0|\chi^{\dagger}\sigma^i\psi| H +\gamma\rangle\langle
H+\gamma|
\psi^{\dagger}(-\frac{i}{2}\tensor{\bf{D}})^2\sigma^i\chi|0\rangle,
\end{eqnarray}
where $F_2({}^3S_1)$ and $G_2({}^3 S_1)$ are the short-distance
coefficients. Again, those short-distance coefficients can be
inferred from perturbative QCD. The long-distance matrix elements
are non-perturbative in nature.

The short-distance coefficients in this case can be determined by
matching the cross section of the free quark pair production process
$e^++e^- \to b\bar{b}(^3S_1)$, with the $b\bar{b}$ pair near the
threshold. The cross section of this process can be evaluated both in
the QCD and in the NRQCD factorization formulism  (\ref{eq:formular-soft}).
The short-distance coefficients can then be determined by
equaling the cross section inferred from both theories. The short-distance
coefficients $F_2({}^3S_1)$ and $G_2({}^3S_1)$ in (\ref{eq:formular-soft}) have been
determined  to the NNLO\cite{A.Czarnecki
1998}\cite{Beneke:1998} and  to
the NLO\cite{Bodwin:2009} in the expansion of $\alpha_s$ in literatures.

\section{  $e^+e^- \to \eta_c +\gamma_{hard}$ \label{sec:3}}
In this section, we apply   NRQCD factorization formulism (\ref{eq:formular-hard}) to the process $e^+e^- \to \eta_c
+\gamma_{hard}$ and determine the short-distance coefficients
$F_1({}^1S_0)$ and $G_1({}^1S_0)$ to the NLO and to the LO in $\alpha_s$
respectively. As mentioned in previous section, they can be determined by matching
the process $e^+(p_{e1}) +e^-(p_{e2}) \to c(p_1)\bar{c}(p_2)(^1S_0)
+ \gamma(k)$ with $c\bar{c}$ in the color-singlet $^1S_0$ state
near the threshold.

Since the $c\bar{c}$ is in the $^1S_0$ state, Lorentz and CPT
invariance implies that the tensor $A_{\mu\nu}$ can only be
expressed as
\begin{eqnarray}
{\cal A}^{\mu\nu}=  A \epsilon_{\mu\nu\alpha\beta} P^\alpha k^\beta
\;, \label{AA}
\end{eqnarray}
where $A$ is a Lorentz invariant quantity and can be inferred from
perturbative QCD. Consequently, the differential and the total  cross section
read:
\begin{eqnarray}
\frac{d\sigma}{d x}
%[e^+e^-\rightarrow c{\bar c}(b{\bar b})({}^1S_0)+\gamma]
&=&\frac{\alpha|A|^2}{64}
(1-\frac{4E^2}{s})^3(1+x^2), \label{sigma-dif}\\
\sigma
%[e^+e^-\rightarrow c{\bar c}(b{\bar b})({}^1S_0)+\gamma]
&=&\frac{\alpha|A|^2}{24}(1-\frac{4E^2}{s})^3\,.\label{sigma-tot}
\end{eqnarray}
Here, we use a notation $x\equiv \cos \theta$, where $\theta$ is the
angle between the photon and the beam line at CM frame.

The cross section for the same process can also be derived
 in the NRQCD factorization formulism. Thus, we can determine the short-distance coefficients by equaling both results.

\subsection{Short-distance coefficients in the LO in $\alpha_s$}

In this subsection, we  determine the short-distance coefficients $
F_1({}^1S_0) $  and $ G_1({}^1S_0) $ to the LO in $\alpha_s$.  They
can be obtained by  matching the cross section of $e^+e^- \to
Q\bar{Q} (^1S_0)  +\gamma $ at tree level both in the QCD and in the NRQCD
factorization formulism. To gain $ G_1({}^1S_0) $ which is the
short-distance coefficient of the relativistic correction  matrix element, we
need to expand the cross section of this process to order ${\rm\bf q}^2$.

Imposing  the on-shell condition of the external quarks on
(\ref{A-mn-tilde}), ${\cal A}^{\mu\nu}$ in (\ref{aaa})  is
simplified as
\begin{eqnarray}\label{A-mn}
{\cal A}^{\mu\nu}=e_Q^2e^2\,{\rm Tr} \,\left[ \,\left(\,
 {\gamma^\nu \not k \gamma^\mu \over 2p_1\cdot k } -
 {\gamma^\mu \not k \gamma^\nu \over 2 p_2\cdot k }
 \right) \, {\Pi}_m \, \right]\;.
\end{eqnarray}
The Lorentz structure of the projection operator $\Pi_0$ can be
written in a general form
\begin{eqnarray}{\label{projector-1-1}}
\Pi_0= a \,\gamma_5 \not\!P + b\, \gamma_5 (\not\!P \not\!q-
\not\!q\not\!P) + c \,\gamma_5\;,
\end{eqnarray}
where $a$, $b$, $c$ are Lorentz invariant
and can be easily inferred from (\ref{projector-1}). One see that
only the first term in (\ref{projector-1-1}) gives contribution to
(\ref{A-mn}). Inserting this term into (\ref{A-mn}), we find
\begin{eqnarray}\label{A-mna}
{\cal A}^{\mu\nu}=-6i\,a\,e_Q^2e^2\,
 \left(
  {1 \over p_1\cdot k } \,+\, {1 \over p_2\cdot k }
  \right)\;
 \epsilon_{\mu\nu\alpha\beta}
P^\alpha k^\beta \;.
\end{eqnarray}
It readily follows that
\begin{eqnarray}\label{A-a}
{ A} =
  {-6i\,a\,e_Q^2e^2\,P\cdot k \over p_1\cdot k \,p_2\cdot k } = {-24i\,a\,e_Q^2e^2\,\over
  P\cdot k} \left( 1+ {{\rm\bf q}^2 \over
 3m^2}   \right) + O({{\rm\bf q}^4 \over m^4})
\;,
\end{eqnarray}
where we have performed a substitution\cite{bodwin-lee2004}
\begin{eqnarray}
 q^\mu q^\nu  \longrightarrow {1\over3}\,\left( \, -g^{\mu\nu} + {P^{\mu}P^{\nu} \over
 4E^2}\right){\rm\bf q}^2 \;.
\end{eqnarray}
Meanwhile, expand $a$ with respect to  ${\rm\bf q}^2$, to obtain
\begin{eqnarray}\label{a}
a= \frac{m}{2\sqrt{6}E}
 =  {1\over 2\sqrt{6} } \, \left(1-{1\over 2}\,{{\rm\bf q}^2 \over
 m^2}\right) + O({{\rm\bf q}^4 \over
 m^4}) \;.
\end{eqnarray}

Combining (\ref{A-a}) and (\ref{a}), we present the total cross section
(\ref{sigma-tot}) in the expansion of ${\rm\bf q}^2$ as:
\begin{eqnarray}
\sigma^{(0)}
 %[e^+e^-\rightarrow c{\bar c}({}^1S_0)+\gamma]
&=& \frac{64e_Q^4\pi^2\alpha^3}{s^2}(1-r)\nonumber\\
 & & -\frac{64e_Q^4\pi^2(1+2r)\alpha^3}{3s^2}\frac{\bf{q}^2}{m^2}+{\cal
O}(\bf{q}^4) \, ,\label{sigma-tot-1}
\end{eqnarray}
where $r \equiv 4m^2/s $ and superscript (0) denotes the
contribution arising from the LO in $\alpha_s$.

Next, we turn to calculate
the production cross section of $e^+e^-\to c\bar{c}({}^1S_0)+\gamma_{hard}$
in the NRQCD factorization formulism (\ref{eq:formular-hard}). At tree level, the matrix elements give:
\begin{eqnarray}
\langle c\bar{c} ({}^1S_0)|\psi^\dagger\chi|0\rangle=2E\sqrt{6},\nonumber\\
\langle c\bar{c} ({}^1S_0)|\psi^\dagger(-\frac{i}{2}\tensor{{\bf
D}})^2\chi|0\rangle=2E\sqrt{6}~{\bf q}^2.
\end{eqnarray}
 Inserting them into
(\ref{eq:formular-hard}) and then expanding (2) in terms of ${\bf q}^2$ up
to order ${\bf q}^2$, we obtain
\begin{eqnarray}\label{eq:formular-hard-2}
\sigma^{(0)} %[e^+e^- \to [c\bar{c}]_1(^1S_0)+\gamma_{hard}]
 &=&
24{F_1^{(0)}({}^1S_0)}(1+{{\bf q}^2 \over m^2} ) + 24
{G_1^{(0)}({}^1 S_0)}{{\bf q}^2 \over m^2} \;.
\end{eqnarray}
By equaling (\ref{eq:formular-hard-2}) with  (\ref{sigma-tot-1}), we determine the
 short distance coefficients $F_1({}^1S_0)$ and
$G_1({}^1S_0)$ to the LO in $\alpha_s$:
\begin{eqnarray}
F_1^{(0)}({}^1S_0)=\frac{8e_Q^4\pi^2\alpha^3}{3s^2}(1-r),\nonumber\\
G_1^{(0)}({}^1S_0)=-\frac{32e_Q^4\pi^2\alpha^3}{9s^2}(1-\frac{r}{4}).
\end{eqnarray}
Finally, the  cross section of $e^+e^-\to \eta_c+\gamma$
  to the LO in $\alpha_s$ and  the  NLO in $v^2$ is given as:
\begin{eqnarray}{\label{nrqcdsigma}}
\sigma^{(0)}&=&\frac{8e_Q^4\pi^2\alpha^3}{3m^2s^2}(1-r)\langle0|\chi^\dagger\psi|\eta_c\rangle\langle\eta_c|\psi^\dagger\chi|0\rangle\nonumber\\
&-&\frac{32e_Q^4\pi^2\alpha^3}{9m^4s^2}(1-\frac{r}{4}){\rm
Re}[\langle0|\chi^\dagger\psi|\eta_c\rangle\langle\eta_c|\psi^\dagger(-\frac{i}{2}\tensor{{\bf
D}})^2\chi|0\rangle].
\end{eqnarray}
Here the two hadron matrix elements are normalized
relativistically, which is related to the standard non-relativistic
normalized hadron matrix elements\cite{chen:1999hg} by
\begin{eqnarray}{\label{transfer}}
\langle0|\chi^\dagger\psi|\eta_c\rangle&=&\sqrt{4m}(\langle0|\chi^\dagger\psi|\eta_c\rangle_{BBL}
+\frac{1}{4m^2}{\rm
Re}[\langle0|\chi^\dagger(-\frac{i}{2}\tensor{{\bf
D}})^2\psi|\eta_c\rangle_{BBL}]),\nn\\
\langle0|\chi^\dagger(-\frac{i}{2}\tensor{{\bf
D}})^2\psi|\eta_c\rangle&=&\sqrt{4m}\langle0|\chi^\dagger(-\frac{i}{2}\tensor{{\bf
D}})^2\psi|\eta_c\rangle_{BBL}.
\end{eqnarray}

Inserting ({\ref{transfer}}) into ({\ref{nrqcdsigma}}), we
obtain the cross section with the hadron matrix elements in
non-relativistic normalization as
\begin{eqnarray}{\label{nrqcdsigma1}}
\sigma^{(0)}&=&\frac{32e_Q^4\pi^2\alpha^3}{3ms^2}(1-r)\langle0|\chi^\dagger\psi|\eta_c\rangle\langle\eta_c|\psi^\dagger\chi|0\rangle_{BBL}\nonumber\\
&-&\frac{16e_Q^4\pi^2\alpha^3}{9m^3s^2}(5+r){\rm
Re}[\langle0|\chi^\dagger\psi|\eta_c\rangle\langle\eta_c|\psi^\dagger(-\frac{i}{2}\tensor{{\bf
D}})^2\chi|0\rangle_{BBL}].
\end{eqnarray}
%%%%%%%%%%%%%%%%%%%%%%%%%%%%%%%%%%%%%%%%%%%%%%%%%%%%%%%%%%%%%%%%%%%%%%%%%%%%%%
\subsection{ The NLO QCD correction \label{subsec-nlo}}
%%%%%%%%%%%%%%%%%%%%%%%%%%%%%%%%%%%%%%%%%%%%%%%%%%%%%%%%%%%%%%%%%%%%%%%%%%%%%%

%%%%%%%%%%%%%%%%%%%%%%%%%%%%%%%%%%%%%%%%%%%%%%%%%%%%%%%%%%%%%%%%%%%%%%%%%%%%%%
\begin{figure}[!hbp]
\centerline{
\includegraphics[width=14.0cm]{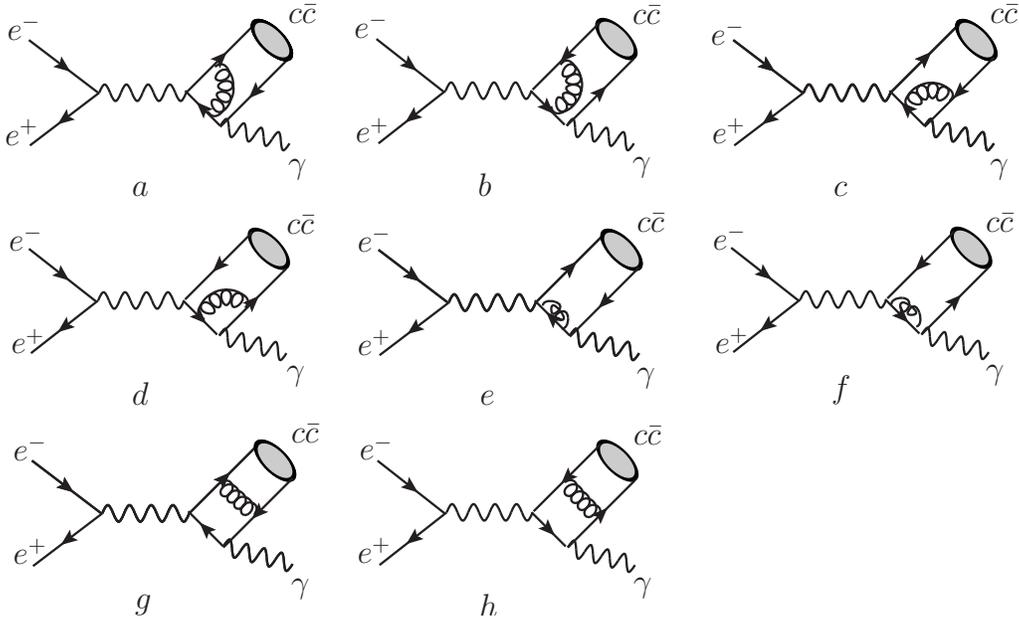}
} \caption{One-loop Feynman diagrams for   $e^+e^- \to c\bar{c}
+\gamma$. } \label{fig2}
\end{figure}
%%%%%%%%%%%%%%%%%%%%%%%%%%%%%%%%%%%%%%%%%%%%%%%%%%%%%%%%%%%%%%%%%%%%%%%%%%%%%%

In this subsection,  we determine the short-distance coefficients
$F_1({}^1S_0)$  to the NLO in $\alpha_s$. To this end, we need to
match the cross section of the process $e^+e^- \to
c\bar{c}(^1S_0)+\gamma_{hard}$ to the one-loop level but the LO in $v^2$ both in the QCD and
in the NRQCD factorization formulism.

%%%%%%%%%%%%%%%%%%%%%%%%%%%%%%%%%%%%%%%%%%%%%%%%%%%%%%%%%%%%%%%%%%%%%%%%

The Feynman diagrams responsible for the one-loop QCD correction are
illustrated in Fig.~{\ref{fig2}}. We perform the renormalization in
on-mass-shell(OS) scheme and take dimensional regularization to
regulate both the ultraviolet(UV) and the infrared(IR)
divergences. In OS scheme,  external quark lines do
not receive any QCD corrections, and the counterterms from the
renormalization constants of the heavy quark wavefunction  and
the heavy quark mass are given by\cite{Chao:2008}:
\begin{eqnarray}
\delta Z_{Q}^{\rm OS}&=&-C_F\frac{\alpha_s}{4\pi}
\left(\frac{1}{\epsilon_{\rm UV}}+\frac{2}{\epsilon_{\rm IR}}
-3\gamma_E+3\ln\frac{4\pi\mu^2}{m_c^2}+4\right)+{\cal
O}(\alpha_s^2), \label{Z-Q}\\
 \label{Z-mQ} \delta Z_{mQ}^{\rm
OS}&=&-C_F\frac{\alpha_s}{4\pi} \left(\frac{3}{\epsilon_{\rm
UV}}-3\gamma_E+3\ln\frac{4\pi\mu^2}{m_c^2}+4\right)+{\cal
O}(\alpha_s^2)\,,
\end{eqnarray}
where $\mu$ is the renormalization scale, $\gamma_E$ is the Euler's
constant, and $C_F=\frac{4}{3}$  for $SU(3)_c$.

% As we did in leading order, first, we calculate the cross section of
% process $e^+e^-\to Q\bar{Q}({}^1S_0)+\gamma$ in pQCD. There are
Among the eight diagrams in Fig.~\ref{fig2}, % to ${\cal O}(\alpha_s)$, which include
there are two self-energy diagrams, four triangle diagrams and two
box diagrams. The divergences of these diagrams can be analyzed as
follows:  the self-energy and the triangle diagrams contain  UV
divergences, while the box diagrams possess IR divergences. In
addition, Coulomb singularities may arise from the box diagrams, due to the exchange
of longitudinal gluon between $c$ and $\bar{c}$.
% we regularized UV divergences and IR divergences in dimension
% regularization and Coulomb singularities by adding a small relative
% velocity $Q$ between the quark pair,
% $v=|\vec{p_Q}-\vec{p}_{\bar{Q}}|/m$, defined in the meson rest
% frame.
Our calculations indicate that the UV divergences from the
self-energy and the triangle diagrams are canceled by that from the renormalization constants
of the quark field and the mass and IR divergences from % the triangle diagrams and
the box diagrams are canceled by that in $\delta Z_Q$.
As one will see, the remaining Coulomb singularities will be canceled by Coulomb
singularities in NRQCD matrix elements by matching condition. In our
practical calculations, the  `Feycalc' package\cite{Feyncalc:1991}
has been employed to take the trace of the $\gamma-$matrix and to
reduce tensor integrals into scalar ones. Furthermore, we carry out
the IR-safe scalar integrals by using G.'t Hooft and M. Veltman's
method\cite{G.thooft:1979}. For the IR-divergent scalar integrals,
we take the results from Ref.~\cite{R.keith Ellis:2007}. After
canceling all these UV and IR divergences, we present the NLO cross
section of $e^+e^-\to c\bar{c}(^1S_0)+\gamma$:
\begin{eqnarray}{\label{includeC}}
\sigma =\sigma^{(0)}\mid_{{\rm\bf q}=0}
\left(1+\frac{\pi^2}{v}C_F\frac{\alpha_s}{\pi}+\frac{\alpha_s}{\pi}C(r)
\right),
\end{eqnarray}
where $\sigma^{(0)}|_{{\rm\bf q}=0}$ is the LO cross section
(\ref{sigma-tot-1}) with  ${\rm\bf q}=0$  and $C(r)$ is a function
 of $r$ and is both UV and IR finite.  The explicit expression for $C(r)$
is given in Appendix~\ref{appendix1}.

Subsequently, we turn to deal with the same process in the NRQCD factorization
formulism. The main ingredient of this calculation is to obtain the  NRQCD matrix element to the NLO in $\alpha_s$, which
reads\cite{Bodwin:1994jh},
\begin{eqnarray}
\vert\langle c\bar{c}({}^1S_0)|\psi^\dagger \chi|0\rangle\vert^2
&=&24m^2\left(1+\frac{\pi^2}{v}C_F\frac{\alpha_s}{\pi}\right),
\end{eqnarray}
where the Coulomb singularity term emerges from the Coulomb gluon
exchanging between the quark and the anti-quark.  The total cross section of $e^+e^-\to c{\bar
c}({}^1S_0)+\gamma$ in the NRQCD factorization formulism can be
readily expressed as
\begin{eqnarray}{\label{NLONRQCD matrixelement}}
\sigma=24F_{1}(^1S_0)\left(1+\frac{\pi^2}{v}C_F\frac{\alpha_s}{\pi}\right).
\end{eqnarray}

By equaling (\ref{includeC}) with
(\ref{NLONRQCD matrixelement}), we determine  the NLO
short-distance coefficient:
\begin{eqnarray}{\label{short-distance coefficient2}}
F_{1}(^1S_0)=F^{(0)}_{1}(^1S_0)\left(1+\frac{\alpha_s}{\pi}C(r)\right),
\end{eqnarray}
from which we see that the Coulomb singularity cancels in the short-distance coefficient as expected.
$1+\frac{\alpha_s}{\pi}C(r)$ is usually referred to the
$K-$factor.

Inserting this coefficient into ({\ref{eq:formular-hard}}),
we reach the final result for the cross section to the NLO QCD correction in $\alpha_s$ and the LO in $v^2$:
\begin{eqnarray}\label{NLOeta}
\sigma=\sigma^{(0)}(1+\frac{\alpha_s}{\pi}C(r)+{\cal
O}(\alpha_s^2)),
\end{eqnarray}
where $\sigma^{(0)}$ is given by (\ref{nrqcdsigma}) with
${\bf q}=0$.

It is instructive to look at the behaviors of $C(r)$ in two limits of
$r\to 0$ and $r\to \infty$. The limit of $r\to 0$ corresponds
to a very high energy production of $\eta_c$. In this limit, $C(r)$ reduces to
\begin{eqnarray}{\label{slimit1}}
\lim_{r\to 0} C(r)
=-\frac{2}{9}\left[(9-6\log{2})\log{r}+9(3+\log^2{2}-3\log{2})+\pi^2\right].
\end{eqnarray}
The coefficient of $\log{r}$ term agrees with that obtained in Ref.~\cite{Jia and
Yang:2009}.

The limit of $r\to \infty$  corresponds to the NLO QCD corrections to the process of
$\eta_c$ decay into two photons. To achieve this limit,  one needs to
recalculate the $K-$factor for $r >1$. We define $C'(r)$ in this
region via
\begin{eqnarray}{\label{etacprime}}
\frac{\sigma(r^\star\to \eta_Q\gamma)}{\sigma^{(0)}(r^\star\to
\eta_Q\gamma)}=1+\frac{\alpha_s}{\pi}C^\prime(r),
\end{eqnarray}
whose expression  is given in Appendix~\ref{appendix1}. With the expression of $C^\prime(r)$ in hand, we take the  limit of $r\to \infty$ for $C^\prime(r)$, which reads
\begin{eqnarray}{\label{limit2}}
\lim_{r\to \infty}C^\prime(r)&=&C_F(-5+\frac{\pi^2}{4}).
\end{eqnarray}
%%%%%%%%%%
It is encouraging that (\ref{limit2}) is consistent with the QCD correction to the cross section of $\eta_c \to
\gamma\gamma$. Comparing our results with that given in
 Ref.~\cite{Shifman:1980dk}, we found that the expressions given in
their Eqs.~(23) (25), from which they obtained
the two limits of $r\to 0$ and $r\to \infty $, were incorrect.
Actually, it is easy to check that the results
given in their Eqs.~(23) (25) go to infinity by taking these two limits. Therefore we claim that
there might be some typos in their results.

%%%%%%%%%%%%%%%

%%%%%%%%%%%%%%%%%%%%%%%%%%%%%%%%%%%%%%%%%%%%%%%%%%%%%%%%%%%%%%%%%%%%%%%%%%%%%%

%%%%%%%%%%%%%%%%%%%%%%%%%%%%%%%%%%%%%%%%%%%%%%%%%%%%%%%%%%%%%%%%%%%%%%%%%%%%%%
%\section{NRQCD formula for a heavy quarkonium production\label{sec:NRQCD formalism}}
%%%%%%%%%%%%%%%%%%%%%%%%%%%%%%%%%%%%%%%%%%%%%%%%%%%%%%%%%%%%%%%%%%%%%%%%%%%%%%

%%%%%%%%%%%%%%%%%%%%%%%%%%%%%%%%%%%%%%%%%%%%%%%%%%%%%%%%%%%%%%%%%%%%%%%%%%%%%%
\section{ $e^+e^- \to \chi_{cJ} + \gamma_{hard}$\label{sec:QCD correction to chi}}
%%%%%%%%%%%%%%%%%%%%%%%%%%%%%%%%%%%%%%%%%%%%%%%%%%%%%%%%%%%%%%%%%%%%%%%%%%%%%%

In this section, we apply  the NRQCD factorization formulism
(\ref{eq:formular-hard}) to the processes $e^+e^- \to
\chi_{cJ} +\gamma_{hard}$ with $J=0,1,2$ and determine the
short-distance coefficients $F_1({}^3P_J)$  to the NLO in $\alpha_s$. As mentioned
in Sec.~\ref{sec:2}, these coefficients can be determined by matching the processes
$e^++e^-\to c\bar{c}(^3P_J) + \gamma$
with the $c\bar{c}$ near the threshold and in the color-singlet
${}^3P_J$ states.

\subsection{The LO short-distance coefficients in $\alpha_s$}
We first determine the LO coefficients  in $\alpha_s$ by calculating
the cross sections of $e^+(p_{e1}) +e^-(p_{e2}) \to
c(p_1)\bar{c}(p_2)(^3P_J) + \gamma(k)$ both in the QCD and in the NRQCD
factorization formulism.  Two tree-level diagrams contributing to the processes
$e^+(p_{e1}) +e^-(p_{e2}) \to c(p_1)\bar{c}(p_2)(^3P_J) + \gamma(k)$
are illustrated in Fig.~\ref{fig1}. To carry out the calculations, we need to construct the
polarization vectors and tensors for $\chi_1$ and $\chi_2$ respectively. Three polarization
vectors $\epsilon^\mu_{\{0,\pm 1 \}}$ for
 $\chi_1$
 can be explicitly
expressed as(provided the hadron moving along $z$ axis.)
\begin{eqnarray}
\epsilon_{+1}^\mu=\frac{1}{\sqrt{2}}(0,-1,-i,0),\ \
\epsilon_{-1}^\mu=\frac{1}{\sqrt{2}}(0,+1,-i,0),\ \
\epsilon_{0}^\mu=\frac{1}{2m}(|{\bf P}|,0,0,E_P).
\end{eqnarray}
Then five polarization tensors for $\chi_2$ are readily written
in terms of $\epsilon^\mu_{\{0,\pm 1\}}$ as
\begin{eqnarray}
\epsilon_{\pm2}^{\mu\nu}&=&\epsilon_{\pm1}^\mu\epsilon_{\pm1}^\nu,\ \ \epsilon_{\pm1}^{\mu\nu}=\frac{1}{\sqrt{2}}(\epsilon_{\pm1}^\mu\epsilon_{0}^\nu+\epsilon_{0}^\mu\epsilon_{\pm1}^\nu),\nonumber\\
\epsilon_{0}^{\mu\nu}&=&\frac{1}{\sqrt{6}}(\epsilon_{+1}^\mu\epsilon_{-1}^\nu+2\epsilon_{0}^\mu\epsilon_{0}^\nu+\epsilon_{-1}^\mu\epsilon_{+1}^\nu).
\end{eqnarray}

The amplitudes of $e^+e^-\to c\bar{c}(^3P_J) + \gamma$  can be
picked out from (\ref{A-mn}) by extracting the contributions corresponding to the $c\bar{c}(^3P_J)$ production. To this end, one needs to expand the amplitudes to
the LO in $q$ and to project it into the $^3P_J$ states with
helicity $\lambda_H$. It can be done by using the
formulae\cite{A.Petrelli:1997}\cite{Braaten:2002fi}:
\begin{eqnarray}{\label{ampformular}}
{\cal A}^{\alpha\beta}(\lambda_H)=|{\bf q}|\frac{\partial}{\partial
q^{\mu}}{\rm Tr}[{\cal \tilde{A}}^{\alpha\beta}\,\Pi_{1\,\nu}] {\cal
P}^{\mu\nu}_J \,,
\end{eqnarray}
where $\Pi^{\nu}_1 $ is defined via $\Pi_1 \equiv \Pi^{\nu}_1
\epsilon^\nu $ and the projection operators
 ${\cal P}_J^{\mu\nu}$ are defined by
\begin{eqnarray}{\label{angularoperator}}
{\cal P}^{\mu\nu}_0&=&\frac{1}{\sqrt{3}}
(-g^{\mu\nu}+\frac{P^{\mu}P^{\nu}}{4E^2}),\nonumber\\
{\cal
P}^{\mu\nu}_1&=&\frac{i}{2\sqrt{2}E}
 \epsilon^{\mu\nu\rho\sigma}P_\rho\epsilon^{*}_{\lambda_H\,\sigma},\nonumber\\
{\cal P}^{\mu\nu}_2&=&\epsilon^{*\mu\nu}_{\lambda_H}.
\end{eqnarray}

With these formulae, it is straightforward to calculate the
cross sections of the processes $e^+e^-\rightarrow c{\bar
c}({}^3P_J(\lambda_H)+\gamma $ with a specific helicity
$\lambda_H$ for $c\bar{c}$. The differential cross sections of these processes in
the LO in $\alpha_s$ yield
\begin{eqnarray}{\label{coefficient1}}
\frac{d\sigma^{(0)}}{dx}=\frac{48e_Q^4\alpha^3\pi^2}{s^2m^2(1-r)}|{
\bf q}|^2\widetilde{F}_J(\lambda_H),
\end{eqnarray}
where the functions $\widetilde{F}_J(\lambda_H)$ are
\begin{eqnarray}
\widetilde{F}_0(0)&=&\frac{(1-3r)^2}{6}(1+x^2),\nonumber\\
\widetilde{F}_1(0)&=&(1+x^2),\nonumber\\
\widetilde{F}_1(\pm1)&=&r(1-x^2),\nonumber\\
\widetilde{F}_2(0)&=&\frac{1}{3}(1+x^2),\nonumber\\
\widetilde{F}_2(\pm1)&=&r(1-x^2),\nonumber\\
\widetilde{F}_2(\pm2)&=&r^2(1+x^2).
\end{eqnarray}

  Since relativistic
corrections of $^3P_J$ production will not be calculated, we have set $M_H=2E=2m$. Our formulae (\ref{coefficient1}) agree with that obtained in
Ref.~\cite{jungil Lee}\cite{Braaten and Chen:1995hg}.

 We now turn to evaluate the cross sections of the same processes in the
NRQCD factorization formulism (\ref{eq:formular-hard}). The matrix elements for $c\bar{c}({}^3P_J)$ production
are given by
\begin{eqnarray}\label{matrix}
 \langle c\bar{c}({}^3P_J(\lambda_H))|\psi^\dagger ({\cal K}_{^3P_J}\cdot \epsilon_J)\chi|0\rangle
 &=&2\sqrt{6} m  |{\bf q}|.
 \end{eqnarray}
%\frac{1}{\sqrt{2}}\langle
%[Q\bar{Q}]_1({}^3P_1)|\psi^\dagger(-\frac{i}{2}\tensor{{\bf
%D}}\times{\bf\sigma}\cdot{\bf\epsilon}_H)\chi|0\rangle &=&2E\sqrt{6}|{\bf q}|,\nonumber\\
%\sum_{ij}\langle
%[Q\bar{Q}]_1({}^3P_2)|\psi^\dagger(-\frac{i}{2}\tensor{D}^{(i}\sigma^{j)}\epsilon^{ij}_H)\chi|0\rangle
%&=&2E\sqrt{6}|{\bf q}|,
%
%\begin{eqnarray}\label{matrix}
%\frac{1}{\sqrt{3}}\langle
%[Q\bar{Q}]_1({}^3P_0)|\psi^\dagger(-\frac{i}{2}\tensor{{\bf
%D}}\cdot{\bf \sigma})\chi|0\rangle &=&2E\sqrt{6}|{\bf q}|,\nonumber\\
%\frac{1}{\sqrt{2}}\langle
%[Q\bar{Q}]_1({}^3P_1)|\psi^\dagger(-\frac{i}{2}\tensor{{\bf
%D}}\times{\bf\sigma}\cdot{\bf\epsilon}_H)\chi|0\rangle &=&2E\sqrt{6}|{\bf q}|,\nonumber\\
%\sum_{ij}\langle
%[Q\bar{Q}]_1({}^3P_2)|\psi^\dagger(-\frac{i}{2}\tensor{D}^{(i}\sigma^{j)}\epsilon^{ij}_H)\chi|0\rangle
%&=&2E\sqrt{6}|{\bf q}|,
%\end{eqnarray}
%the factor of $2E$ appearing on the right sides of
%Eq.(\ref{matrix}), is due to that the states
%$|[Q\bar{Q}]_1({}^1S_0)\rangle$ on the left sides are normalized
%relativistically. Immediately,
Thus, the differential cross sections over $x$ within the framework of NRQCD
factorization formulism read
\begin{eqnarray}
\frac{d\sigma^{(0)}}{dx}=24F_1^{(0)}({}^3P_J(\lambda_H))|{\bf q}|^2,
\end{eqnarray}
where $F_1^{(0)}({}^3P_J(\lambda_H))$ denote the LO short-distance
coefficients. By equaling the cross sections on the QCD side with that on the
NRQCD side, we determine the short-distance coefficients:
\begin{eqnarray}
F_1^{(0)}({}^3P_J(\lambda_H))=\frac{2e_Q^4\alpha^3\pi^2}{m^2s^2(1-r)}\widetilde{F}_J(\lambda_H).
\end{eqnarray}

 Consequently, the differential cross sections of $e^+e^-\to
\chi_{cJ}(\lambda_H)\gamma$ in the LO in $v$ and $\alpha_s$ are
written as:
\begin{eqnarray}\label{abc}
\frac{d\sigma^{(0)}}{dx}=\frac{F_1^{(0)}({}^3P_J)}{m^2}\langle0|\chi^\dagger{\cal
K}_J\psi|\chi_J(\lambda_H)\rangle\langle\chi_J(\lambda_H)|\psi^{\dagger}{\cal
K}_J\chi|0\rangle,
\end{eqnarray}
where  the $\chi_{J}$ states in the matrix element are normalized
relativistically. To express our formulae as the product of
short-distance coefficients and standard non-relativistic
normalized matrix elements, we can use the relations between
relativistic normalized matrix elements and non-relativistic normalized
matrix elements\cite{chen:1999hg}
\begin{eqnarray}
\langle0|\chi^\dagger{\cal
K}_{^3P_J}\psi|H\rangle=\sqrt{2M_H}\langle0|\chi^\dagger{\cal
K}_{^3P_J}\psi|H\rangle_{BBL},
\end{eqnarray}
where $|H\rangle_{BBL}$ is the non-relativistic normalized matrix element
used in Ref.~\cite{Bodwin:1994jh} and $M_H$ is the mass of $H$. In the
LO in $v$, we set $M_H=2E|_{q=0}=2m$.

%With the normalization of the {\it BBL} matrix-element by
%Eq.~(\ref{matrix element change}),
The differential cross sections of $e^+e^- \to
\chi_{cJ}(\lambda_H)+\gamma$ can be reexpressed as:
\begin{eqnarray}
\frac{d\sigma^{(0)}}{dx}
=\frac{8e_Q^4\alpha^3\pi^2}{m^3s^2(1-r)}\widetilde{F}_J(\lambda_H)\langle0|\chi^\dagger{\cal
K}_J\psi|\chi_{cJ}\rangle\langle\chi_{cJ}(\lambda_H)|\psi^{\dagger}{\cal
K}_J\chi(\lambda_H)|0\rangle_{BBL}.
\end{eqnarray}

The total cross sections are achieved by integrating over $x$ and
summing over all the helicity states $\lambda_H$ of the $\chi_J$.
We present the final results:
\begin{eqnarray}\label{LOchi}
\sigma^{(0)}( \chi_{cJ}) &=& {F}_J^{(0)} \langle0|\chi^\dagger{\cal
K}_J\psi|\chi_{cJ}\rangle\langle\chi_{cJ}|\psi^{\dagger}{\cal
K}_J\chi|0\rangle_{BBL} \, ,
\end{eqnarray}
where ${F}^{(0)}_J $ for $J=0,1,2$  are given by
\begin{eqnarray}\label{F-J-0}
{F}^{(0)}_0 &=&\frac{32e_Q^4\alpha^3\pi^2(1-3r)^2}{9m^3s^2(1-r)},\nn\\
{F}^{(0)}_1&=&\frac{64e_Q^4\alpha^3\pi^2(1+r)}{3m^3s^2(1-r)},
,\nn\\
{F}^{(0)}_2&=&\frac{64e_Q^4\alpha^3\pi^2(1+3r+6r^2)}{9m^3s^2(1-r)}.
\end{eqnarray}

\subsection{ Short-distance coefficients in the NLO in $\alpha_s$}
In this subsection, we determine  the short-distance coefficients in
the NRQCD factorization formulism for processes $e^+e^- \to
\chi_{cJ}+\gamma$ up to the NLO in $\alpha_s$. As done in the
case of $\eta_c$ production, we calculate the cross sections of the
processes $e^+e^- \to c\bar{c}(^3P_J) +\gamma$ both in the
QCD and in the NRQCD factorization formulism to one-loop level. Feynman diagrams responsible
for the processes are the same as that of $^1S_0$ production
as illustrated in Fig.~\ref{fig2}. We still carry the renormalization in
on-mass-shell(OS) scheme. The wave function renormalization constant
and mass renormalization constant have been given in
(\ref{Z-Q})(\ref{Z-mQ}).

The UV and IR divergences  of each diagram are similar with that
of $\eta_c$ production. The self-energy and triangle diagrams
contain ultraviolet(UV) divergences. The box diagrams have both
infrared(IR) divergences and Coulomb singularities. Combining
contributions from the eight diagrams and the counterterms, all divergences
cancel except Coulomb singularities which will be canceled by
Coulomb singularities in NRQCD matrix elements.
The cross sections of the processes $e^+e^-\rightarrow c{\bar
c}({}^3P_J,\lambda_H) + \gamma $ in the NLO in $\alpha_s$ are given by
\begin{eqnarray}
\frac{d\sigma}{dx}&=&\frac{48e_Q^4\alpha^3\pi^2}{m^2s^2(1-r)}|{ \bf
q}|^2\widetilde{F}_J(\lambda_H)\left(1+\frac{\pi^2}{v}C_F
\frac{\alpha_s}{\pi} %\nonumber\\&+&
 +\frac{\alpha_s}{\pi}C_J^{\lambda_H}(r)\right),
\end{eqnarray}
where the index $\lambda_H$  denotes the helicity of $\chi_{cJ}$, $\widetilde{F}_J(\lambda_H)$ have been given in previous
subsection and the expressions for $C_J^{\lambda_H}(r)$ are listed in
Appendix~\ref{appendix2}.

One can immediately get the total cross sections by integrating
over
 $x$ and summing over the polarizations of $\chi_{cJ}$. As a
result, the cross sections yield
\begin{eqnarray}{\label{pwaveqcd1}}
\sigma &=& F^{(0)}_{J} \vert{\bf q}\vert^2
\,\left(1+\frac{\pi^2}{v}C_F\frac{\alpha_s}{\pi}
 +\frac{\alpha_s}{\pi} C_J(r)\right)\,
 \end{eqnarray}
where $C_J(r)$ for $J=0,1,2$ are given by
\begin{eqnarray}{\label{pwaveqcd}}
  C_0(r) &=& C_0^0(r),\nn\\
 C_1(r) &=&\frac{C_1^0(r)+rC_1^1(r)}{1+r},\nn\\
 C_2(r) &=&\frac{C_2^0(r)+3rC_2^1(r)+6r^2C_2^2(r)}{1+3r+6r^2}.
\end{eqnarray}

We now  calculate the cross sections of the same processes in the
NRQCD factorization formulism. The NRQCD matrix elements
of $P$-wave operators in the LO in $v$ and NLO in $\alpha_s$ are
given by
\begin{eqnarray}
\vert\langle c\bar{c}({}^3P_0)|{\cal K}_{^3P_J}|0\rangle\vert^2
&=&24m^2|{\bf
q}|^2\left(1+\frac{\pi^2}{v}C_F\frac{\alpha_s}{\pi}\right).
\end{eqnarray}
With it, one readily obtains  the cross sections of
$e^+e^-\rightarrow c{\bar c}({}^3P_J)+\gamma$ on the NRQCD side:
\begin{eqnarray}{\label{pwavenrqcd}}
\sigma &=&24F_1({}^3P_J)|{ \bf
q}|^2[1+\frac{\pi^2}{v}C_F\frac{\alpha_s}{\pi}].
\end{eqnarray}

The short-distance coefficients $F_1({}^3P_J)$ can be determined by
equaling (\ref{pwaveqcd1}) with (\ref{pwavenrqcd}),
\begin{eqnarray}
F_1({}^3P_J)&=&F_1^{(0)}({}^3P_J)[1+\frac{\alpha_s}{\pi}C_J(r)].
\end{eqnarray}

Finally, the cross sections of the processes
$e^+e^-\rightarrow \chi_{cJ}+\gamma_{hard} $  in the NLO in $\alpha_s$ can be expressed as
\begin{eqnarray}{\label{NLOchi}}
\sigma(\chi_{cJ})&=&
\sigma^{(0)}(\chi_{cJ})\left(1+\frac{\alpha_s}{\pi}C_J(r)\right),
\end{eqnarray}
where $\sigma^{(0)}_{cJ}$ denote the LO cross sections given in
(\ref{LOchi}).

It is interesting to look at the behaviors of $C_J(r)$ in two limits of
$r\to 0 $ and $r\to \infty$. The limit of $r\to 0$ is related to the
$\chi_{cJ}$ production in a very high energy. In this limit, we find
\begin{eqnarray}{\label{plimit1}}
\lim_{r\to 0}C_0(r)&=&-\frac{2}{9} \left[(3-6 \log2) \log r+9\log^2
2-33
   \log2+\pi ^2\right],\nn\\
\lim_{r\to 0} C_1(r) &=&-\frac{2}{9} \left[(9-6 \log2) \log
r+9\log^2 2-15 \log2+\pi ^2+21\right],\nn\\
\lim_{r\to 0} C_2(r) &=&-\frac{2}{9} \left[(3-6\log2) \log r+9
\log^2 2+3\log2+\pi
   ^2+18\right].
\end{eqnarray}
%%%%%%%%%%%%%%
The limit of $r\to \infty$ is related to the NLO QCD corrections to the cross sections of $\chi_{cJ} \to \gamma\gamma$. As in the case of $\eta_c$, one needs to recalculate the $K-$factor for $r>
1$. We define $C'_{cJ}(r)$ via
\begin{eqnarray}
\frac{\sigma(r^\star\to \chi_J\gamma)}{\sigma^{0}(r^\star\to
\chi_J\gamma)}=1+\frac{\alpha_s}{\pi}C_J^\prime(r).
\end{eqnarray}
Explicit expressions of  $C^{\prime}_{cJ}(r)$ are listed in
Appendix~\ref{appendix2}. In the limit of $r\to \infty$, $C^\prime_{J}(r)$ go
to:
\begin{eqnarray}{\label{plimit2}}
\lim_{r\to \infty}C_0^\prime &=&C_F(-\frac{7}{3}+\frac{\pi^2}{4}),\nn\\
 \lim_{r\to \infty}C_2^\prime &=&-4C_F.
\end{eqnarray}
The formulae(\ref{plimit2}) are desirable, since these
are nothing but the QCD corrections to the processes
$\chi_{0,2}\to \gamma\gamma$. In (\ref{plimit2}), we did not
take the limit of $r\to \infty$ for $C^\prime_{1}(r)$, since the cross section of $\chi_{c1}$ decay
into $\gamma\gamma$ vanishes due to the Yang's theory.

\section{  $e^+e^- \to \eta_b(\chi_{bJ})+\gamma_{soft}$ \label{sec:4}}

As discussed in Sec. II, for bottomonium production at the B
factories through process $e^+ +e^- \to \,{\rm bottomonium  }\, +
\gamma $, the emitted photon is soft comparing to the $b$ quark
mass. The  cross section of the process is expressed as
(\ref{eq:formular-soft}) in the framework of the NRQCD factorization
formulism. The short-distance coefficients $F_2({}^3S_1)$ and
$G_2({}^3S_1)$ in (\ref{eq:formular-soft}) can be determined by
matching the cross section of the process $e^+(p_{e1}) +e^-(p_{e2})
\to c(p_1)\bar{c}(p_2) $ with the $c\bar{c}$ in the color-singlet
$^3S_1$ state predicted by QCD and by the NRQCD factorization formulism.
It has been calculated up to order  $\alpha_s^2$. Here we list only
the $F_2({}^3S_1)$ result\cite{Bodwin:2002hg}\cite{A.Czarnecki 1998}
\cite{Beneke:1998} \cite{Barbieri:1975}\cite{W.Celmaster:1979}:
\begin{eqnarray}
F_2({}^3S_1) &=&\frac{4\pi^3
e_Q^2\alpha^2\delta{(s-4m^2)}}{m}\left\{1-4C_F\frac{\alpha_s}{\pi}+\left[-117.46+0.82n_f+\frac{140\pi^2}{27}\log{\frac{2m}{\mu}}\right]\frac{\alpha_s^2}{\pi^2}
\right\}\nn\\
\end{eqnarray}
The  matrix elements in (\ref{eq:formular-soft}) describe
probabilities of a point-like $^3S_1$ $b\bar{b}$ state transiting
into the $^1S_0$ or the $^3P_J$ state through a magnetic $M1$
transition or an electric $E1$ transition by emitting one soft
photon. Although they are QED processes, the matrix elements are
non-perturbative in nature. One may evaluate those $M1$ and $E1$
transition processes  by perturbative QCD  and relate the matrix
elements appearing in (\ref{eq:formular-soft})  to those  appearing
in (\ref{eq:formular-hard}). However, this kind of calculations is
not reliable. Some additional Coulomb singularities may arise from
gluon exchange between almost on-shell $b$ and $\bar{b}$ quarks.
Terms like $\alpha_s/v$ appearing in radiative corrections spoil
fixed order perturbation calculations on those matrix elements. Even
at tree level,  it is proportional to a factor $s/4m^2 -1$  for the
$\eta_b$ production and $1/(s/4m^2 -1)$   for the $\eta_{cJ} $
production. Thus the results are very sensitive to the quark mass.
To gain a reliable prediction on the matrix in
(\ref{eq:formular-soft}), one need to invoke some nonperturbative
methods such as  lattice QCD.

%%%%%%%%%%%%%%%%%%%%%%%%%%%%%%%%%%%%%%%%%%%%%%%%%%%%%%%%%%%%%%%%%%%%%%%%%%%%%%
\section{Numerical results and discussion\label{sec:numerical results}}
%%%%%%%%%%%%%%%%%%%%%%%%%%%%%%%%%%%%%%%%%%%%%%%%%%%%%%%%%%%%%%%%%%%%%%%%%%%%%%
In this section, we carry our numerical calculations on the
production cross sections of $e^+e^- \to H \,+\gamma\,(H=
\eta_c,\eta_c(2S),\chi_{cJ}) $ at the B factories  using the results
derived in  section III and section IV. To this end, we need to
determine some relevant parameters such as quark mass,
$\alpha_s(\mu)$, and those NRQCD matrix elements. We will estimate
them below.

\subsection{Input parameters}{\label{sec:input parameters}}
At the B factories, we take $\sqrt{s}=10.6 {\rm GeV}$ and
$\alpha(\mu=\sqrt{s})=1/131$. The quark mass is taken as $m_c=1.4
\textrm{GeV}$. There is an ambiguity in choosing the value of $\mu$
in $\alpha_s$ since there are several energy scales involved, such as
$\sqrt{s}$ and heavy quark mass. In the limit of $r\to 0$,
one can use the leading log approximation (LLA) to resume those
large logarithmic terms such as $(\alpha_s(\sqrt{s}) \ln \sqrt{s}/m
)^n$\cite{Shifman:1980dk,Jia and Yang:2009}. This accounts for the
effects of the running of the strong coupling constant $\alpha_s$
from energy scale $\sqrt{s}$ to $m$. However, in doing this
approximation, one has omitted constant terms and terms suppressed
by  power of $r$. To recover these contributions, one needs to carry
out calculations to higher loops and to include contributions from
higher dimensional operators. Those  are much more complicated. In
our full one-loop results, we keep all these terms but leave an
ambiguity in choosing the value of $\mu$ in $\alpha_s(\mu)$. To
estimate the uncertainty caused by it, we take three different
values of $\mu$. They are $\sqrt{s}$, $2m_c$, and $ m_c$. The
corresponding values of $\alpha_s(\mu)$ are
$\alpha_s(\sqrt{s})=0.17$, $\alpha_s(2m_c)=0.24$, and
$\alpha_s(m_c)=0.30$.

%Before estimating the cross section and QCD radiative corrections,
%we need some input parameters, which consist of the masses of the
%heavy quarks, the center of mass frame energy $\sqrt{s}$, the QED
%and QCD coupling constant $\alpha$, $\alpha_s$ and the hadron matrix
%elements. In the numerical estimation, we will take $m_c=1.4 GeV$,
%$m_b=4.6 GeV$, $\sqrt{s}=10.6 GeV$, and
%$\alpha(\mu=\sqrt{s})=1/131$. In addition, we will take the QCD
%coupling constants $\alpha_s$ in three different scales to estimate
%the QCD corrections, for it's hard to know the exact scale of the
%strong coupling. That is $\alpha_s(\sqrt{s})=0.17$,
%$\alpha_s(2m_c)=0.24$, $\alpha_s(m_c)=0.30$.
 %for charmonium
 %and $\alpha_s(\sqrt{s})=0.17$, $\alpha_s(2m_b)=0.18$,
 %$\alpha_s(m_b)=0.21$ for bottomonium.

 As for the hadron matrix
elements, we will take them as same as those used in  \cite{jungil
Lee}\cite{bodwin:0710hg}\cite{bodwin:2006},
\begin{eqnarray}\label{numericalmatrix}
\langle0|\chi^{\dagger}\psi|\eta_c\rangle\langle\eta_c|\psi^\dagger\chi|0\rangle&=&0.437\textrm{GeV}^3,\nonumber\\
\langle0|\chi^{\dagger}\psi|\eta_c(2S)\rangle\langle\eta_c(2S)|\psi^\dagger\chi|0\rangle&=&0.274\textrm{GeV}^3,\nonumber\\
 \frac{1}{3}\vert\langle
\chi_{c0}|\psi^\dagger(-\frac{i}{2}\tensor{{\bf D}}\cdot{\bf
\sigma})\chi|0\rangle\vert^2 &=&0.051 \textrm{GeV}^5 \nonumber\\
\frac{1}{2}\vert\langle
\chi_{c1}|\psi^\dagger(-\frac{i}{2}\tensor{{\bf
D}}\times{\bf\sigma}\cdot{\bf\epsilon}_H)\chi|0\rangle\vert^2 &=&0.060 \textrm{GeV}^5,\nonumber\\
\vert\sum_{ij}\langle
\chi_{c2}|\psi^\dagger(-\frac{i}{2}\tensor{D}^{(i}\sigma^{j)}\epsilon^{ij}_H)\chi|0\rangle\vert^2
&=&0.068 \textrm{GeV}^5.
 %,\nonumber\\
 %\frac{1}{3}\vert\langle
 %\chi_{b0}|\psi^\dagger(-\frac{i}{2}\tensor{{\bf D}}\cdot{\bf
 %\sigma})\chi|0\rangle\vert^2 &=&2.03 GeV^5, \nonumber\\
 %\frac{1}{2}\vert\langle
 %\chi_{b1}|\psi^\dagger(-\frac{i}{2}\tensor{{\bf
 %D}}\times{\bf\sigma}\cdot{\bf\epsilon}_H)\chi|0\rangle\vert^2 &=&2.03 GeV^5,\nonumber\\
 %\vert\sum_{ij}\langle
 %\chi_{b2}|\psi^\dagger(-\frac{i}{2}\tensor{D}^{(i}\sigma^{j)}\epsilon^{ij}_H)\chi|0\rangle\vert^2
 %&=&2.03 GeV^5.
\end{eqnarray}

For the $\eta_c$ production,  in order to assess the
contribution from the relativistic correction,  we will
evaluate the NRQCD matrix element corresponding to the relativistic
correction by using the Gremm-Kapustin\cite{Gremm:1999hg} relation,
\begin{eqnarray}{\label{G-K relationvelocity}}
{\rm Re}\langle0|\chi^{\dagger}(-\frac{i}{2}\tensor{\bf{D}})^2
\psi|\eta_c\rangle\langle\eta_c|\psi^\dagger\chi|0\rangle&=&
m^2v^2\, {\rm Re}\langle0|\chi^{\dagger}
\psi|\eta_c\rangle\langle\eta_c|\psi^\dagger\chi|0\rangle .
\end{eqnarray}
We take $v^2=0.13$ for $\eta_c$ as that in Ref.~\cite{Sang:2009} and
the correspondent NLO NRQCD matrix element yields:
\begin{eqnarray}
{\rm
Re}\langle0|\chi^{\dagger}(-\frac{i}{2}\tensor{\bf{D}})^2\psi|\eta_c\rangle\langle\eta_c|\psi^\dagger\chi|0\rangle&=&0.111\textrm{GeV}^5.
\end{eqnarray}

For the $\eta_c(2S)$ production,  the LO NRQCD matrix element,
which is decided by fitting data from $\Gamma[\psi(2S)\to e^+e^-]$ based on heavy-quark
spin symmetry, is accurate only in the LO in $v$ as pointed out in
Ref.~\cite{jungil Lee}. Therefore we will not consider the relativistic correction to $\eta_c(2S)$ production for consistence.

 In the following subsection, we will use these input parameters
to calculate the production cross sections.
%%%%%%%%%%%%%%%%%%%%%%%%%%%%%%%%%%%%%%%%%%%%%%%%%%%%%%%%%%%%%%%%%%%%%%%%%%%%%%

\subsection{Numerical calculations}
%%%%%%%%%%%%%%%%%%%%%%%%%%%%%%%%%%%%%%%%%%%%%%%%%%%%%%%%%%%%%%%%%%%%%%%%%%%%%%
With the parameters given above, we can evaluate the cross sections
for the production of $\eta_c$, $\eta_c(2S)$, and $\chi_{cJ}$ by
using Eqs.~(\ref{nrqcdsigma}), (\ref{NLOeta}), (\ref{LOchi}), and
 (\ref{NLOchi}). For comparison, we list the results of the LO and the
NLO in $\alpha_s$ in Table~\ref{table1}. For the NLO results, we
have taken $\mu$ in $\alpha_s(\mu)$ to be $\sqrt{s}$, $2m_c$, and
$m_c$. In Tabel~\ref{table1}, we have omitted  relativistic
corrections. From the table, we see that most of the QCD radiative
corrections are negative and considerable. For some processes like
$\chi_2$ production, the corrections are unexpectedly large.

Then we can estimate the relativistic corrections by
Eq.~(\ref{nrqcdsigma}). After accounting for the NLO radiative QCD
correction(we take $\mu=2m_c$ in $\alpha_s(\mu)$) and the
relativistic correction, the total cross section of the $\eta_c$
production decrease from 83.3fb to 58.2fb, where the NLO radiative
QCD correction and relativistic correction contribute -15.3fb and
-9.81fb respectively. We see that both of them give considerably
negative contributions.

\begin{table}[t]
\caption{\label{table1} Predicted cross sections (in fb)  for the
production of $\eta_c$, $\eta_c(2S)$, and $\chi_{cJ}$ with various
$\mu$. ${\sigma^{(0)}}$ and  $ {\sigma}$ are the cross sections with
the LO and the NLO short-distance coefficients, repectively. $
{\sigma \over \sigma^{(0)}}-1$ is relative corrections caused by the
NLO corrections.}
\begin{ruledtabular}
\begin{tabular}{lc|cc|cc|cc}
 \multicolumn{2}{c}{}& \multicolumn{2}{c} {$ \mu=\sqrt{s} $} &
  \multicolumn{2}{c} {$ \mu=2m $} &
 \multicolumn{2}{c} {$ \mu= m $} \\
\hline& {$\sigma^{(0)}$}  & {$\frac{\sigma}{\sigma^{(0)}}-1$}
&{$\sigma$}& {$\frac{\sigma}{\sigma^{(0)}}-1$} & {$\sigma$}&
{$\frac{\sigma}{\sigma^{(0)}}-1$}&
   {$\sigma$}\\
%  & \multicolumn{1}{c} {$\sigma^{(0)}(fb)$}  &
%   \multicolumn{1}{c}{$\frac{\sigma}{\sigma^{(0)}}-1$} &
%   \multicolumn{1}{c}{$\sigma$}&
%   \multicolumn{1}{c}{$\frac{\sigma^{(1)}}{\sigma^{(0)}}(\alpha_s(2m))$} &
%   \multicolumn{1}{c}{$\sigma^{(total)}(fb)$}&
%   \multicolumn{1}{c}{$\frac{\sigma^{(1)}}{\sigma^{(0)}}(\alpha_s(m))$}&
%   \multicolumn{1}{c}{$\sigma^{(total)}(fb)$}\\
\hline $\eta_{c}$&83.3 & -13.0\% & 72.5 & -18.4\% & 68.0 & -23.0\% & 64.1 \\
\hline $\eta_{c}(2S)$&52.2 & -13.0\% & 45.4 & -18.4\% & 42.6 & -23.0\% & 40.2 \\
%\hline $\eta_{b}$&2.9 & -23.8\% & 2.21 & -25.2\% & 2.17 & -29.4\% & 2.05 \\
\hline $\chi_{c0}$&1.19 & 10.4\% & 1.31 & 14.7\% & 1.36 & 18.4\% & 1.41 \\
\hline $\chi_{c1}$&14.3 & -17.1\% & 11.9 & -24.1\% & 10.9 & -30.1\% & 10.0 \\
\hline $\chi_{c2}$&6.28 & -48.9\% & 3.21 & -69.0\% & 1.95 & -86.2\% & 0.87 \\
%\hline $\chi_{b0}$&0.80 & -42.9\% & 0.46 & -45.5\% & 0.44 & -53.1\% & 0.38 \\
%\hline $\chi_{b1}$&5.28 & -34.3\% & 3.47 & -36.3\% & 3.36 & -42.3\% & 3.05 \\
%\hline $\chi_{b2}$&6.70 & -37.5\% & 4.19 & -39.7\% & 4.04 & -46.4\% & 3.59 \\
\end{tabular}
\end{ruledtabular}
\end{table}

%%%%%%%%%%%%%%%%%%%%%%%%%%%%%%%%%%%%%%%%%%%%%%%%%%%%%%%%%%%%%%%%%%%%%%%%%%%%%%

These theoretical predictions for the cross sections receive
uncertainties from the ambiguities of the values of input
parameters. These parameters are the $c$ quark pole mass, the NRQCD
matrix elements, and the $\mu$ in  $\alpha_s(\mu)$. The $c$ quark
pole mass may effect both the short-distance coefficients and NRQCD
matrix elements. Especially, the matrix element of the relativistic
correction term is very sensitive to the value of the pole mass
when the Gremm-Kapustin\cite{Gremm:1999hg} relation (\ref{G-K
relationvelocity}) is used to estimate its value. For more details,
we refer the reader to Ref.~\cite{jungil Lee}. Another theoretical
uncertainties arise from the higher order QCD corrections and higher
order relativistic corrections. Including these corrections, the
$\mu$ dependence and uncertainties of the matrix elements may be
reduced. Varying the values of all these parameters in  reasonable
ranges and omitting higher order correction terms both in $\alpha_s$
and in $v$, we can estimate theoretical uncertainties on the
predicted cross sections. The results are listed in
Table~\ref{table2}.

\begin{table}[t]
\caption{\label{table2} { Predicted  cross sections (in  fb) for the
production of $\eta_c$, $\eta_c(2S)$, $\chi_{cJ}$ by including
theoretical uncertainties.}}
\begin{ruledtabular}
\begin{tabular}{cccccc}

%  & \multicolumn{1}{c} {$\sigma^{(0)}(fb)$}  &
%   \multicolumn{1}{c}{$\frac{\sigma}{\sigma^{(0)}}-1$} &
%   \multicolumn{1}{c}{$\sigma$}&
%   \multicolumn{1}{c}{$\frac{\sigma^{(1)}}{\sigma^{(0)}}(\alpha_s(2m))$} &
%   \multicolumn{1}{c}{$\sigma^{(total)}(fb)$}&
%   \multicolumn{1}{c}{$\frac{\sigma^{(1)}}{\sigma^{(0)}}(\alpha_s(m))$}&
%   \multicolumn{1}{c}{$\sigma^{(total)}(fb)$}\\
 &$\eta_{c}$& $\eta_c(2S)$ & $\chi_{c0}$ & $\chi_{c1}$ & $\chi_{c2}$ \\
\hline  & $68.0^{+22.2}_{-20.3}$& $42.6^{+10.9}_{-8.8}$ & $1.36^{+0.26}_{-0.26}$&$10.9^{+3.7}_{-3.4}$ &$1.95^{+1.85}_{-1.56}$ \\
\end{tabular}
\end{ruledtabular}
\end{table}

In $e^+e^-$ collisions, the heavy quarkonium with $C-$parity even
can also be produced via the double photon processes. At very high
energy limit, this process can be described by  the equivalent
photon approximation(EPA). Comparing to the processes $e^+e^- \to H
\,+\gamma\,(H= \eta_c,\eta_c(2S),\chi_{cJ}) $,  the total production
cross sections of the double photon processes are suppressed by
additional power of $\alpha$ but enhanced by
$\log^2{s/m_e^2}$\cite{Brodsky:1970}. The cross sections of the $s-$
channel processes $e^+e^- \to H \,+\gamma\,(H=
\eta_c,\eta_c(2S),\chi_{cJ}) $ are suppressed by $M_h^4/s^2$.
Combining all these factors, at the B factories, the production
cross sections of  the heavy quarkonium with $C-$parity even via the
double photon process are larger than that via associated production
with photon. However, the signals in detectors from these two
production mechanisms are very different. Especially, the produced
heavy quarkonia via double photon processes are predominated by small
$P_T$ events at high energy which are hard to be detected from their
decay products. On the other hand, the produced heavy quarkonia via
associated production with photon are insensitive to the $P_T$. Moreover, the
events can be reconstructed by measuring the photon assuming the photon
detector is good enough. A thorough analysis on these processes in
the framework of the NRQCD factorization formalism will be presented
elsewhere.

\begin{acknowledgments}
After this work was finished, we were told that D.~Li,  Z.G.~He, and
K.T.~Chao had done a similar work\cite{He:2009hg}. One of the authors,
Wen-Long~Sang would like to thank G.~T.~Bodwin and Yu~Jia for useful
discussion and to thank Zhi-Guo~He for checking results. This work
was supported partly by the National Natural Science Foundation of
China (NNSFC) under No.10875156.
\end{acknowledgments}
%%%%%%%%%%%%%%%%%%%%%%%%%%%%%%%%%%%%%%%%%%%%%%%%%%%%%%%%%%%%%%%%%%%%%%%%%%%%%%
%%%%%%%%%%%%%%%%%%%%%%%%%%%%%%%%%%%%%%%%%%%%%%%%%%%%%%%%%%%%%%%%%%%%%%%%%%%%%%
\appendix
\section{The NLO radiative corrections to short distance coefficients for
the $\eta_{c}$ production}{\label{appendix1}} The definition for
$C(r)$ appeared in Eq.~(\ref{includeC}):
\begin{eqnarray}{\label{cetaQ}}
C(r)&=&-\frac{2[30r^2-(84+\pi^2)r+2\pi^2+54]}{9(2-r)(1-r)}+\frac{8(2r-3)}{3(2-r)^2}\log(\frac{2}{r}-2)-\frac{4}{b}\log(\frac{1-b}{1+b})\nn\\
%-----------------------------------
&+&\frac{2}{3(r-1)}[(1+\frac{r}{2})\log^2(\frac{1-b}{1+b})-\log^2(\frac{2}{r}-1)]+\frac{4}{3(1-r)}{\rm
Li_2}(\frac{r}{2-r}),
\end{eqnarray}
where we use $r=\frac{4m^2}{s}$, $b=\sqrt{1-4m^2/s}$ and $m=m_c$ in
Eq.~(\ref{includea1}). In addition, dilogarithm(${\rm Li}_2(x)$) is
used in the Appendices~\ref{appendix1} and \ref{appendix2}, which is
defined as
\begin{eqnarray}
{\rm Li_2}\equiv-\int^x_0\frac{dz}{z}\log(1-z),
\end{eqnarray}
where $x\le1$.

%%%%%%%%%%%%%%%%%%%%%%%%%%%%%%%%%%%%%%%%%%%%%%%%%%%%%%%%%%%%%%%%%%%%%%%%%%%%%%
The definition for $C^\prime$(r) appeared in Eq.~(\ref{etacprime}):
\begin{eqnarray}{\label{cprimeetaQ}}
C^\prime(r)&=&C_F\frac{\pi^2(r-2)(3r+4)-12(r-1)(5r-9)}{12(r-1)(r-2)}+C_F\frac{2(2r-3)}{(r-2)^2}\log(2-\frac{2}{r})\nn\\
&+&\frac{C_F}{r-1}\{{\rm
Li_2}(\frac{2}{r}-1)-\tan^{-1}{(\frac{1}{\sqrt{r-1}})}[6\sqrt{(r-1)}\nn\\
&+&(r+2)\tan^{-1}{(\frac{1}{\sqrt{r-1}})}]\},
\end{eqnarray}

\section{The NLO radiative corrections to short distance coefficients for
the  $\chi_{J}$ production}{\label{appendix2}}
\subsection{$\chi_{0}$}
\begin{eqnarray}{\label{includea1}}
C_0^0(r)&=&\frac{-2}{9\left(3 r^3-r^2+3 r-1\right)}\left[\frac{\pi
^2(4 r^4-7 r^3+11
r^2-r+1)}{(1-r)^2}\right.\nn\\
&+&\left.\frac{6r\left(7 r^4-25 r^3+31 r^2-25 r+24\right)}{(2-r)^2}\right]\nn\\
%-----------------------------------
&+&\frac{1}{3
(r-1)^2(3r-1)}\left\{\frac{2(4r^4-7r^3+11r^2-r+1)\log^2(\frac{2}{r}-1)}{1+r^2}\right.\nn\\
%-----------------------------------
&+&(3r^3-6r^2-3r+2)\log^2(\frac{1-b}{1+b})+\frac{4(1-r)}{b}(3r^2-12r+5)\log(\frac{1-b}{1+b})\nn\\
%-----------------------------------
&-&\frac{8(2r^4-5r^3+6r^2-2r+1)\log(\frac{2}{r}-2)\log(\frac{2}{r}-1)}{(1+r^2)}\nn\\
%-----------------------------------
&+&\left.\frac{8(8r^5-50r^4+125r^3-152r^2+87r-18)\log(\frac{2}{r}-2)}{(r-2)^3}\right\}.\nn\\
%-----------------------------------
&+&\frac{4 \left[\left(4 r^4-7 r^3+11
   r^2-r+1\right) \text{Li}_2\left(\frac{r}{2-r}\right)-2 \left(2 r^4-5 r^3+6 r^2-2 r+1\right)
   \text{Li}_2\left(2-\frac{2}{r}\right)\right]}{3
   (3r-1)(1-r)^2(1+r^2)}.
\end{eqnarray}
\subsection{$\chi_1$}
\begin{eqnarray}{\label{C_i^1}}
C_1^{\pm1}(r)&=&\frac{\pi ^2 \left(-4 r^2+19 r-5\right)-36
(r-1)^2}{9 (r-1)^2}+\frac{\left(-3 r^2+3 r-2\right) \log
   ^2\left(\frac{1-b}{1+b}\right)}{3(r-1)^2}\nn\\
%-----------------------------------
&+&\frac{4[(r-2)(2r^2-r+1)\log
   \left(\frac{1-b}{1+b}\right)+b \left(2 r^2-5 r+1\right) \log
   \left(\frac{2}{r}-2\right)]}{3b(r-2)(r-1)}\nn\\
   &+&\frac{[\left(4 r^2-19 r+5\right) \log \left(\frac{2}{r}-1\right)-2
   \left(r^2-16 r+3\right) \log
   \left(\frac{2}{r}-2\right)]\log \left(\frac{2}{r}-1\right)}{3(r-1)^2}\nn\\
%-----------------------------------
&+&\frac{2 \left(4 r^2-19 r+5\right)
   \text{Li}_2\left(\frac{r}{2-r}\right)-2 \left(r^2-16 r+3\right)
   \text{Li}_2\left(2-\frac{2}{r}\right)}{3 (r-1)^2}.\nn
\end{eqnarray}

\begin{eqnarray}
C_1^0(r)&=&\frac{2[\pi ^2(r-2)(r^2-r+2)-6(r-1)(4r-7)]}{9(r-1)(r-2)}
+\frac{4(r^2-2r-1)\log
   \left(\frac{1-b}{1+b}\right)}{3b(1-r)}\nn\\
%-----------------------------------
&+&\frac{8(3r^3-10r^2+9r-1)\log(\frac{2}{r}-2)}{3(r-2)^2(r-1)}+\frac{(r-2)(r^2+1)\log^2(\frac{1-b}{1+b})}{3 (r-1)^2}\nn\\
%-----------------------------------
&-&\frac{2 \log \left(\frac{2}{r}-1\right)
   [2(1-2r) \log(\frac{2}{r}-2)+(r-1)(r^2-r+2)\log(\frac{2}{r}-1)]}{3
   (r-1)^2}\nn\\
%-----------------------------------
&-&\frac{4[(1-2 r)
   {\rm Li}_2(2-\frac{2}{r})+(r-1)(r^2-r+2){\rm Li}_2(\frac{r}{2-r})]}{3
   (r-1)^2}.
\end{eqnarray}
\subsection{$\chi_2$}
\begin{eqnarray}
C_2^{\pm2}(r)&=&-\frac{2 r^2[24 (r-1)^2+\pi ^2 (r+3)]}{9
(r-1)^2}+\frac{8r^2[4 b \log \left(\frac{2}{r}-2\right)-(r-5) \log
   \left(\frac{1-b}{b+1}\right)]}{3b(r-1)}\nn\\
%-----------------------------------
   &+&\frac{2r^2\{\log \left(\frac{2}{r}-1\right)[(r+3) \log
   \left(\frac{2}{r}-1\right)-2 (r-1) \log
   \left(\frac{2}{r}-2\right)]-4 \log
   ^2\left(\frac{1-b}{b+1}\right)\}}{3 (r-1)^2}\nn\\
%-----------------------------------
&+&\frac{4 r^2[(r+3) \text{Li}_2\left(\frac{r}{2-r}\right)-(r-1)
   \text{Li}_2\left(2-\frac{2}{r}\right)]}{3 (r-1)^2},\nn
\end{eqnarray}
\begin{eqnarray}
C_2^{\pm1}(r)&=&\frac{\pi ^2(2-r)(2r^2-20r+3)-12r(r-1)^2}{9(r-2)
   (r-1)^2}+\frac{\left(3 r^2+21 r-2\right) \log
   ^2\left(\frac{1-b}{b+1}\right)}{3
   (r-1)^2}\nn\\
%-----------------------------------
   &+&\frac{\log \left(\frac{2}{r}-1\right)[\left(2 r^2-20 r+3\right) \log \left(\frac{2}{r}-1\right)-2
   \left(5 r^2+r+1\right) \log \left(\frac{2}{r}-2\right)]}{3
   (r-1)^2}\nn\\
%-----------------------------------
   &-&\frac{4[(19 r+3)(r-2)^2\log(\frac{1-b}{1+b})+b(10 r^3-49 r^2+51 r+10) \log(\frac{2}{r}-2)]}{3b(r-2)^2(r-1)}\nn\\
%-----------------------------------
&-&\frac{2[\left(5 r^2+r+1\right)
   \text{Li}_2\left(2-\frac{2}{r}\right)-(2 r^2-20 r+3)
   \text{Li}_2\left(\frac{r}{2-r}\right)]}{3 (r-1)^2},\nn
\end{eqnarray}
\begin{eqnarray}
C_2^0(r)&=&-\frac{\pi ^2 \left(2 r^4+9 r^3+27 r^2-11 r+5\right)
(r-2)^2+12
   (r-1)^2 \left(2 r^3-11 r^2+6 r+12\right)}{9 (r-2)^2 (r-1)^2}\nn\\
%-----------------------------------
   &-&\frac{\left(3
   r^3+30 r^2+3 r+2\right) \log ^2\left(\frac{1-b}{1+b}\right)}{3
   (r-1)^2}+\frac{\left(2 r^4+9 r^3+27 r^2-11
   r+5\right) \log^2(\frac{2}{r}-1)}{3
   (r-1)^2}\nn\\
%-----------------------------------
   &+&\frac{8(6r^5-38r^4+60r^3+16r^2-69r+6) \log
   \left(\frac{2}{r}-2\right)}{3(r-2)^3(r-1)}+\frac{4(21r^2+18r-1)\log(\frac{1-b}{1+b})}{3b(r-1)}\nn\\
%-----------------------------------
   &-&\frac{2 \left(2 r^4+9 r^3-9
   r^2-11 r+3\right) \log \left(\frac{2}{r}-2\right)\log
   \left(\frac{2}{r}-1\right)}{3
   (r-1)^2}\nn\\
%-----------------------------------
&+&\frac{2}{3
   (r-1)^2}[(-2 r^4-9 r^3+9 r^2+11 r-3)
   \text{Li}_2(2-\frac{2}{r})\nn\\
&+&(2 r^4+9 r^3+27 r^2-11
   r+5) \text{Li}_2(\frac{r}{2-r})].
\end{eqnarray}
\subsection{the expressions for $C_J^\prime(r)$}
For $C_0^\prime(r)$:
\begin{eqnarray}
C_0^\prime(r)&=&\frac{C_F}{6 (r-2)^2 (r-1)^2(3r-1)}[-6r(7 r^2-25r+24)(r-1)^2\nn\\
&+&\pi^2(r-2)^2(3r-1)-6\pi(r-2)^2\sqrt{r-1}(3r^2-12r+5)]\nn\\
&+&\frac{2
C_F(2r-3)(4r^3-15r^2+19r-6)\log(2-\frac{2}{r})}{(r-2)^3(r-1)(3r-1)}-\frac{C_F
   \text{Li}_2\left(\frac{2}{r}-1\right)}{(r-1)^2}\nn\\
&+&\frac{C_F}{(r-1)^2(3r-1)}\tan^{-1}(\sqrt{r-1})[2\sqrt{r-1}(3r^2-12r+5)\nn\\
&+&\pi(3 r^3-6 r^2-3 r+2)-(3 r^3-6 r^2-3 r+2) \tan
   ^{-1}\left(\sqrt{r-1}\right)].\nn\\
\end{eqnarray}
For $C_1^\prime(r)$:
\begin{eqnarray}
C_1^\prime(r)&=&\frac{C_F}{12(r-2)(r-1)^2(r+1)}[-12(r-1)^2(3r^2-2r-7)\nn\\
&+&\pi
   ^2(r-2)(r^3+r^2+2)+12\pi\sqrt{r-1}(r-2)(2r^3-2r^2+3r+1)]\nn\\
&+&\frac{C_F \left(2 r^4-3 r^3-9 r^2+16 r-2\right) \log
   \left(2-\frac{2}{r}\right)}{(r-2)^2 \left(r^2-1\right)}-\frac{C_F
   \left(r^3+r^2+2\right) \text{Li}_2\left(\frac{2}{r}-1\right)}{2
   (r-1)^2 (r+1)}\nn\\
&+&\frac{C_F}{(r-1)^2(r+1)}\tan^{-1}(\sqrt{r-1})[-2\sqrt{r-1}(2r^3-2r^2+3r+1)\nn\\
&-&\pi(2 r^3-r^2+r+2)+(2r^3-r^2+r+2)\tan^{-1}(\sqrt{r-1})].
\end{eqnarray}
For $C_2^\prime(r)$:
\begin{eqnarray}
C_2^\prime(r)&=&\frac{-C_F}{12 (r-2)^2 (r-1)^2(6r^2+3
   r+1)}[12 \left(24 r^4-91 r^3+79 r^2+6
r+12\right)
   (r-1)^2\nn\\
   &+&\pi ^2 (r-2)^2(9 r^3-21 r^2-6 r-2)+12\pi\sqrt{r-1}(r-2)^2(12r^3-24r^2-9r+1)]\nn\\
&+&\frac{C_F \left(30 r^5-157 r^4+249 r^3-76 r^2-78 r+12\right) \log
   \left(2-\frac{2}{r}\right)}{(r-2)^3(r-1)(6r^2+3r+1)}\nn\\
&+&\frac{C_F \left(9 r^3-21 r^2-6 r-2\right)
   \text{Li}_2\left(\frac{2}{r}-1\right)}{2 (r-1)^2 \left(6 r^2+3
   r+1\right)}\nn\\
&-&\frac{C_F}{(r-1)^2(6 r^2+3
   r+1)}\tan^{-1}(\sqrt{r-1})[-2\sqrt{r-1}(12r^3-24r^2-9r+1)\nn\\
&+&\pi(-6 r^3+15 r^2+9
   r+2)+(6 r^3-15 r^2-9 r-2) \tan
   ^{-1}(\sqrt{r-1})]\nn\\
\end{eqnarray}

%---------------------------------------------------------------------------

\end{document}